\documentclass[aip, apl, reprint,preprintnumbers, nofootinbib, amsmath]{revtex4-2}

\usepackage[]{graphicx}
\usepackage{epstopdf}
\usepackage{bm}
\usepackage{MnSymbol}
\usepackage{latexsym}
\usepackage{amsmath}
\usepackage{refstyle}
\usepackage{hyperref}
\usepackage[utf8]{inputenc}
\usepackage{float}
\usepackage{times}
\usepackage{xcolor}
\usepackage{ulem}
\usepackage{soul}

\begin{document}

\title{Carrier thermalization and zero-point bandgap renormalization in halide perovskites from the Urbach tails of the emission spectrum}
\author{Kingshuk Mukhuti}\thanks{Authors to whom correspondence should be addressed: kingshukmukhuti.phy@gmail.com and bhavtosh@iiserkol.ac.in}
\affiliation{Department of Physical Sciences, Indian Institute of Science Education and Research Kolkata, Mohanpur, Nadia 741246, West Bengal, India}
\author{Arnab Mandal}
\affiliation{Department of Chemical Sciences and Centre for Advanced Functional Materials, Indian Institute of Science Education and Research Kolkata, Mohanpur, Nadia 741246, West Bengal, India}
\author{Basabendra Roy}
\affiliation{Department of Physical Sciences, Indian Institute of Science Education and Research Kolkata, Mohanpur, Nadia 741246, West Bengal, India}
\author{Sayan Bhattacharyya}
\affiliation{Department of Chemical Sciences and Centre for Advanced Functional Materials, Indian Institute of Science Education and Research Kolkata, Mohanpur, Nadia 741246, West Bengal, India}
\author{Bhavtosh Bansal}\thanks{Authors to whom correspondence should be addressed: kingshukmukhuti.phy@gmail.com and bhavtosh@iiserkol.ac.in}
\affiliation{Department of Physical Sciences, Indian Institute of Science Education and Research Kolkata, Mohanpur, Nadia 741246, West Bengal, India}

\begin{abstract}
We develop techniques to study the temperature dependent localization, thermalization, and the effects of phonon scattering on the excitons in halide perovskites from the analysis of the emission spectra.
The excitonic Urbach edge, when inferred from the low energy tails of the temperature dependent luminescence spectra, is shown to be sensitive to the electron distribution and thermalization. A method to observe the Urbach focus is devised for halide perovskites where the temperature dependence of the excitonic gap is anomalous. The value of the zero-point bandgap renormalization is inferred to be about 33 meV. This small value of the bandgap renormalization rules out the formation of small polarons and points to weak electron-phonon coupling. The experiments are performed on the nanosheets of the archetypal halide perovskite, CsPbBr$_3$.
\end{abstract}
\maketitle
Halide perovskites \cite{Yamada2021, Dey2021, Jiang2019, Tao2021, Wu2021, Lao2020,Lao2018, Shibata2020,Sebastian2015} with ABX$_3$ composition (A = Cs or CH$_3$NH$_3$, etc., B = Pb or Sn, and X = Cl, Br, or I) are direct gap semiconductors. Their physical properties are well-described by the standard band theory and the excitons in these systems are of the Mott-Wannier type, albeit with somewhat larger binding energy than GaAs or InP \cite{Miyata2015, Yang2018}. Yet, while conventional semiconductors are sensitive to parts-per-million-level imperfections and require stringent clean-room and controlled growth conditions,  Pb-based halide perovskites (HP) can be prepared via the solution route and still exhibit the excellent light emission characteristics at room temperature.\cite{Umebayashi2003, Kang2017} Thus the HPs combine some of the most desirable characteristics of the inorganic (high mobility) and the organic semiconductors (simplicity of preparation). But beyond the observation of strong light emission at room temperature, one must also understand the effect of defects on nature of the carrier distribution and the role of phonon scattering for optimal materials and device design.\cite{Yamada2021, Sebastian2015, Ledinsky2019, Caselli2020, Li2017}

In this Letter, we have extended the available techniques to examine the complete emission lineshape to uncover hitherto unutilized information regarding the carrier thermalization and the strength of electron-phonon interaction. In particular, we have tailored the Urbach edge \cite{Kost,Dow-Redfield, Schafer-Wegener,Cohen_Economou_John,Cody,Johnson-Tiedje, Falsini2022} analysis so that the luminescence, rather than the absorption spectra, can be studied within this framework. We also reformulate the analysis to accommodate the anomalous increase of the bandgap with temperature in these materials\cite{Kingshuk_APL} and demonstrate the full Urbach rule,\cite{Basab_JPhys, Rupak_PRL} including the Urbach focus,  which we identify with the zero-temperature unrenormalized energy gap.\cite{Cody,Johnson-Tiedje,Basab_JPhys,Rupak_PRL} Experiments are done on  the archetypal halide perovskite, CsPbBr$_3$,  prepared in the form of two-dimensional nanosheets.\cite{Lao2020,Lao2018}

\begin{figure}[h]
	\includegraphics[scale=0.30]{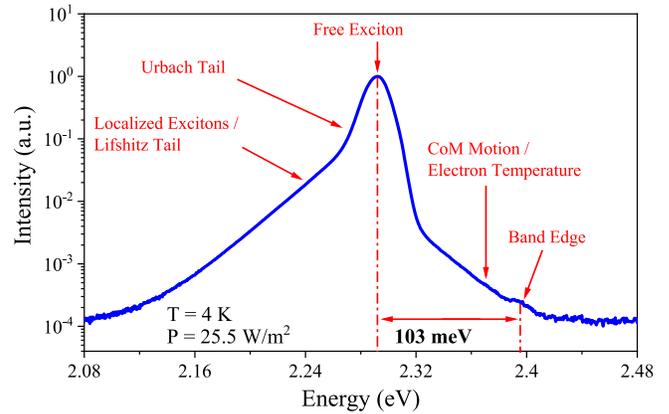}
	\centering
	\caption{Representative PL spectrum from an ensemble of CsPbBr$_3$ nanosheets, plotted on the semi-logarithmic scale. Different regions of the emission line can be attributed to different recombination mechanisms, starting with the localized Lifshitz tail states at the lowest energy to features of the free exciton luminescence (the Urbach tail, the exciton line and the thermalized tail), and finally the band edge. The free exciton binding strength of 103 meV is calculated from the energy difference between the free exciton resonance line to the band edge.}
	\label{Fig:1}
\end{figure}

Figure $\ref{Fig:1}$ shows the $4$ K photoluminescence (PL) spectrum measured at the relatively high laser excitation power ($\approx 25$ Wcm$^{-2}$) where the features from the localized states are minimized. Apart from the prominent free exciton peak at $E_0=2.292$ eV, based on the two-dimensional correlation analysis, the feature at 2.275 eV is assigned to a bound exciton peak which evolves into the exponential Urbach tail  of the free exciton emission (see Supplementary Material Figure S2).

Finally, the Urbach tail itself continuously merges into other localized (the Lifshitz tail \cite{Ziman}) states with a stretched exponential form, extending to $\approx 100$ meV below the free exciton peak. On the high energy side, the band edge at $E_g=2.395$ eV yields the free exciton binding energy $E_b\approx 103$ meV. Considering that the emission is sustained till room temperature, this value for the binding energy seems reasonable (see Supplementary Material).

\begin{figure}[h]
	\includegraphics[scale=0.3]{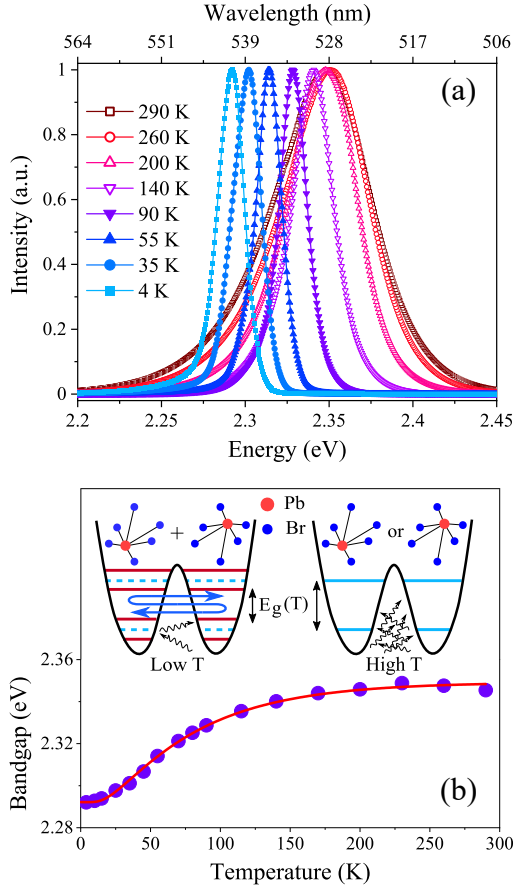}
	\centering
	\caption{(a) A few representative PL spectra plotted at different temperatures, on a linear scale. (b) The anomalous blue shift with temperature is attributed to emphanisis (the basic idea is pictorially represented in the inset, following Ref. \onlinecite{Kingshuk_APL}), and can be fitted to the three parameter formula [Equation $\ref{Eq:5}$]. }
	\label{Fig:2}
\end{figure}

We mainly focus on the low energy tail of the free exciton absorption line inferred from the PL emission spectrum. The region of interest is where the absorption coefficient $\alpha(\hbar\omega,T)$ is $\sim 10^{-1}-10^{-2}$ of its peak value and is characterized by the nearly universal Urbach rule, viz.,
\begin{equation}\label{Eq:1}
	\alpha(\hbar\omega,T)=\alpha_{0}\exp\left[{(\hbar \omega-\varepsilon_F)/E_U(T)}\right].
\end{equation}
$\varepsilon_F$ is called the Urbach focus and $\alpha_{0}$ is the value of the absorption coefficient at $\varepsilon_F$. The slope $E_U(T)$ is also observed to have universal temperature dependence \cite{Dow-Redfield, Toyozawa_Book}
\begin{equation}\label{Eq:2}
	E_U(T)=E_U^{0}\coth\frac{\hbar\Omega_{p}}{2k_{B}T}.
\end{equation}
Here $\Omega_p$ is the characteristic phonon frequency and can be the weighted contributions of many modes and branches.\cite{Basab_JPhys}

In conventional insulators, the existence of a temperature-independent Urbach focus at energy $\varepsilon_F$ is fundamentally dependent on the fact that the temperature dependent shift of the excitonic band edge $E_g(T)$ (or indeed, any of the band structure critical points) is also well-approximated by the same equation\cite{Basab_JPhys}
\begin{equation}\label{Eq:3}
	E_g(T)=\varepsilon_g^0- \Delta E_g^0\coth\frac{\hbar\Omega_{p}}{2k_{B}T}.
\end{equation}
This suggests that the Urbach slope $E_U(T)$ and $E_g(T)$ are both governed by the same exciton-phonon interactions, whose strengths would scale as the Bose distribution  ($n_X (\Omega_p,T)$) that controls the phonon population; $\coth{\hbar\Omega_{p}\over 2k_BT}= 2 [n_X (\Omega_p,T)+{1\over 2}]$, with an added factor of one-half from the zero-point effects, since even at zero temperature, the atoms in a real crystal are dynamic entities with a zero point motion that contributes to the bandgap renormalization.\cite{Basab_JPhys, Rupak_PRL, Antonius2015}

The similarity of Equation $\ref {Eq:2}$ and $\ref{Eq:3}$ implies that the functional form of the Eq. $\ref{Eq:1}$ is invariant with respect to a transformation
\begin{equation}\label{Eq:4}
	\varepsilon_F\rightarrow \tilde{\varepsilon}_F(T)=\varepsilon_F-\Delta E^0_g\coth\left[{\hbar\Omega_{p}\over 2k_{B}T}\right],
\end{equation}
if we also simultaneously redefine $\alpha_{0}\rightarrow \tilde{\alpha}_{0}=\alpha_{0}\exp[E^0_g/E^0_U]$. The existence of the Urbach focus $\varepsilon_F$, the point where the band tails at different temperatures all meet, is thus the happy consequence of the fact that the temperature-dependent decrease of the energy gap and the temperature dependence of the Urbach slope are governed by the same phonon physics. Furthermore, due to the above mentioned factor of ${1\over 2}$, the Urbach tails have remarkable manifestations of the zero-point effects of the electron-phonon interactions, much like the Lamb shift physics in the context of quantum electrodynamics.\cite{Basab_JPhys, Rupak_PRL} The Urbach focus $\varepsilon_F$ denotes the value of unrenormalized energy gap, that is, the bandgap one would hypothetically measure if the electron-phonon interactions were switched off. $\lim_{T\rightarrow 0} E_U(T)=E_U^0 \neq 0$ is the zero-temperature broadening from the electric field of the electron-polar optical phonon vacuum.

Figure $\ref{Fig:2}$ (a) shows the PL spectra at a few representative temperatures. This sigmoidal-shaped [Figure $\ref{Fig:2}$ (b)] anomalous blue-shift of the excitonic gap with temperature is characteristic of all the lead- and tin-containing halide-perovskites.\cite{Sebastian2015}  The primary cause for the change in the excitonic gap in these materials is not in the usual temperature-dependent thermal expansion and enhanced electron-phonon scattering, but rather a high-temperature local symmetry-breaking phenomenon (termed {\it emphanisis}) resulting from the lead atoms' lone-pair stereochemistry.\cite{Fabini2016} Such a variation of temperature-dependent excitonic bandgap can be very well described by semi-empirical expression that is motivated by the physics of the dissipative tunneling of these Pb atoms within the perovskite cage.\cite{Kingshuk_APL}
The exciton gap $E_g(T)$ may be fitted to the three-parameter formula\cite{Kingshuk_APL}
\begin{equation}\label{Eq:5}
	E_g(T)=E_g^\infty-\Xi_0 \rm exp[-\{\rm exp(T_0/T)-1\}^{-1}].
\end{equation}
Here, $E_g^\infty$ is the bandgap at high temperature, $T_0$ is the characteristic phonon temperature, and $|\Xi_0|$ is the bandgap change due to emphanisis. Solid line in Figure $\ref{Fig:2} $ (b) shows the fit of Equation $\ref{Eq:5}$ to the experimental excitonic bandgap data in CsPbBr$_3$, and the parameter values obtained by the fit are $E_g^\infty = 2.349$ eV, $\Xi_0 = 57$ meV, and $T_0 = 62$ K.
\begin{figure}[h]
	\includegraphics[scale=0.3]{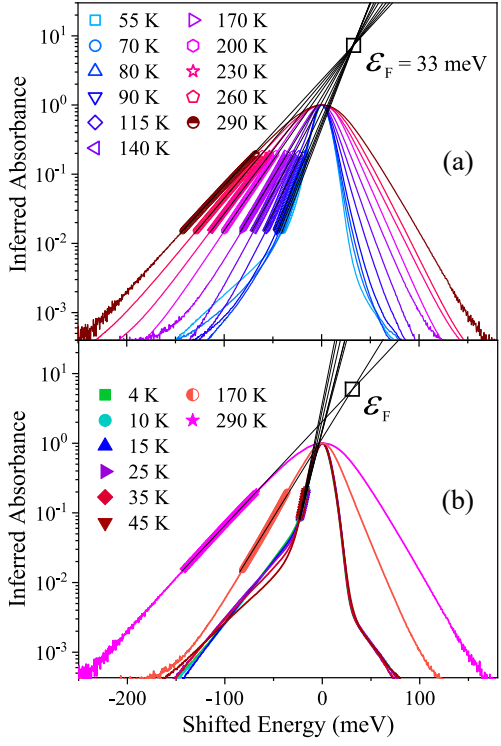}
	\centering
	\caption{Absorption spectra inferred from the PL spectra at different temperatures. (a) At higher temperatures ($>$ 50 K), the Urbach tails converge at a focus. (b) At low temperatures ($<$ 50 K), the tails have completely opposite slopes, and there is no existence of a well-defined focus. This suggests that carriers trapped at low temperatures.}
	\label{Fig:3}
\end{figure}
As the temperature dependence of the gap in these materials has an extraneous cause and Equation \ref{Eq:3} is not valid, $\varepsilon_F$ in Equation $\ref{Eq:1}$ would not be temperature independent and the Urbach rule would not be valid. The Urbach focus analysis\cite{Rupak_PRL} can still be salvaged nevertheless if we go back a step and correct for the emphanisis-induced bandgap shift by translating the peaks to zero energy using  Equation $\ref{Eq:5}$. We have done this in Figure $\ref{Fig:3}$ by fitting straight lines to the inferred absorption data in the selected range ($\sim 10^{-1}-10^{-2}$ of its peak value) and extrapolating them to get an estimate for  $\varepsilon_F$, which  is identified with the zero temperature bandgap renormalization.\cite{Basab_JPhys} The relatively small value (33 meV) of zero-point renormalization suggests that we have Mott-Wannier excitons with weak exciton-phonon coupling. Furthermore, the Urbach energy $E_U$ at any particular temperature is thus the inverse of the slope of the fitted straight line in Figure $\ref{Fig:3}$.

\begin{figure}[h]
	\includegraphics[scale=0.192]{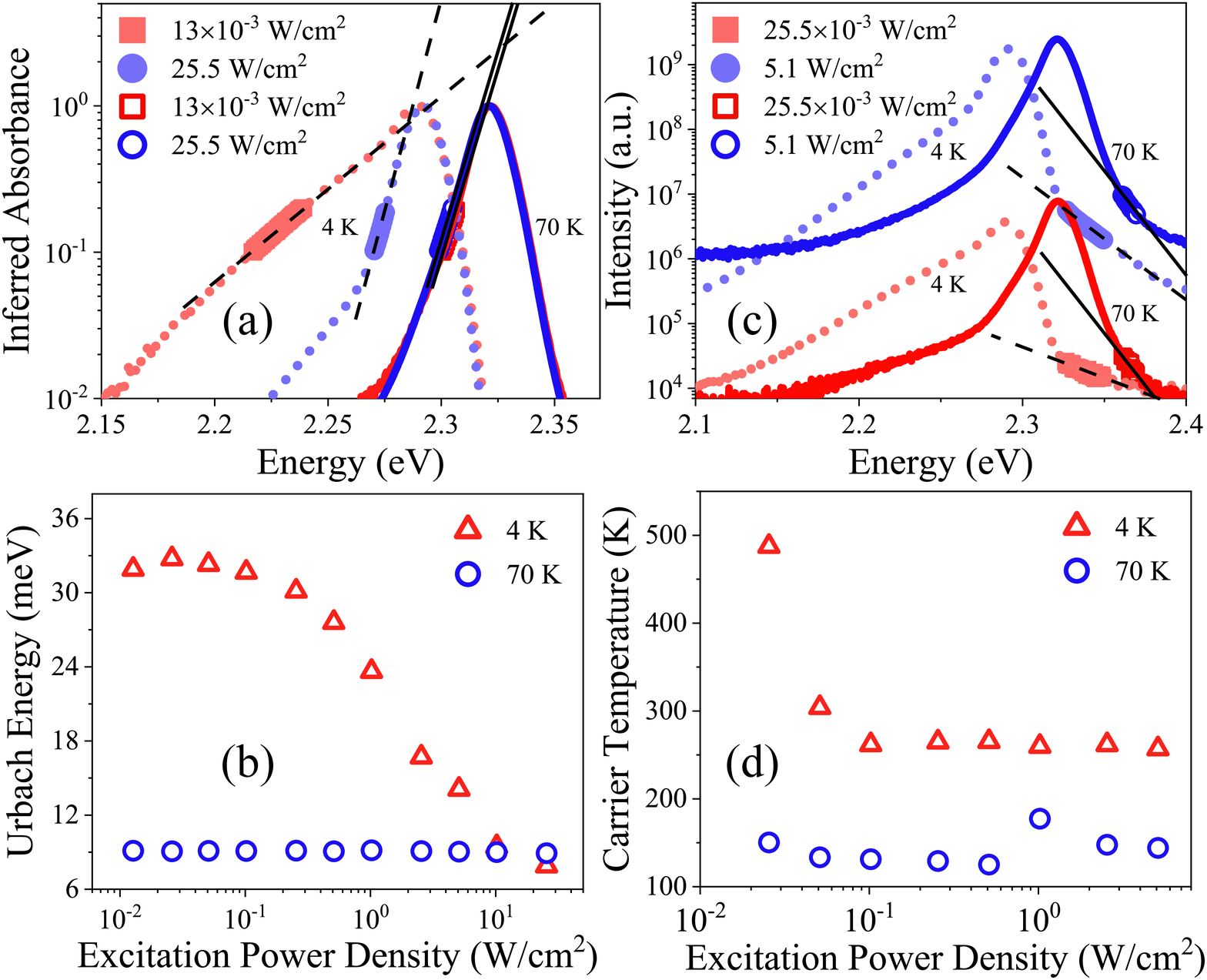}
	\centering
	\caption{Excitation density dependence of the Urbach tails and carrier temperature shows that defect-induced localization is important at 4 K but at 70 K carriers are nearly thermalized.  (a) Slope of the Urbach tails are independent of the excitation powers above 70 K, but changes considerably at 4 K. (b) Calculated Urbach energies at 4 K and 70 K as a function of excitation density. (c) Slope of the exponential high energy tails are nearly independent above 70 K. At 4 K, the slopes strongly depends on excitation density. (d) Carrier temperature as a function of excitation power density at 4 K and 70 K.}
	\label{Fig:4}
\end{figure}

The PL and the absorption phenomena are proportional to the Einstein $A$ and $B$ coefficients and one may thus infer a relationship between PL and absorption measurements via the fundamental connection between the $A$ and $B$.\cite{Rupak_APL} The optical absorption measurement (in the limit of low excitation) can usually be considered to be the measurement of the imaginary component of the sample's susceptibility that is itself proportional to the optical joint density of states. The PL spectrum $PL(\hbar\omega)$ may thus be written as the combination of the joint density of states $\kappa(\hbar\omega)$ and the carrier distribution function, viz.,
\begin{equation}\label{Eq:6}
	PL(\hbar\omega)\propto \kappa(\hbar\omega) n(E-E_0, \mu, T).
\end{equation}
For the $\nu=1$ excitonic emission, where $\nu$ is the principal quantum number, $\kappa(\hbar\omega)$ is a (broadened) resonance line starting at $E_0=E_g-E_b$ that, on the high energy side, evolves with the $\sqrt{E}$ density of hydrogenic states with finite center-of-mass momentum $\vec{K}$, merging with the closely spaced $\nu=2$ and higher bound states leading up to the gap. While optical transitions with $\vec{K}\neq 0$ are usually considered forbidden, note that small relaxations in the vicinity of $\vec{K}= 0$ are not quite ruled out on account of various scattering process, especially at higher temperatures. The emission between the bandedge and the free exciton peak is a region ($2.31-2.40$ eV) attributed to such $\vec{K}\neq 0$ states. The exponential high energy tail is identified with the distribution function and it can be used to infer an approximate electron temperature by fitting the slope to $e^{-E/k_BT_X}$, where $k_B$ is the Boltzmann constant and $T_X$ the inferred carrier temperature [see Figure $\ref{Fig:4}$ \textcolor{red}{(d)}] \cite{Pettinari}. Due to the exponential nature of this high energy tail of the distribution function, the effect of the (polynomial) variation of the density of states with energy can be ignored.

Emission spectrum involves the carrier distribution. Since the total number of carriers participating in the emission process is many orders of magnitude smaller than number of states within the band, the emission spectrum is extremely sensitive to the disorder landscape. The localization of the excitons at these trapping sites prevents their spatial and spectral diffusion necessary for them to attain an equilibrium distribution.
Even though the disorder-localized states may be a negligible fraction of the extended band states, these are selectively occupied by the photoexcited carriers, especially at low temperatures. The low temperature PL is largely representative of these localized states\cite{Peters2021} and thus very sensitive to excitation power.  An increased carrier density resulting from high excitation power would fill these low lying localized states and act to smoothen the potential energy landscape, thereby effectively reducing the disorder. The absorption spectrum, on the other hand, is overwhelmed by the much larger number of extended band states. At the lowest temperatures ($T \simeq 20 $ K), the emission spectrum therefore follows the density of the localized states where the excitons are first trapped. As the temperature is increased, partial thermalization may be accomplished. The low temperature PL spectra, therefore, show various anomalous features which yield direct information about the nature of disorder-induced potential energy landscape localizing the excitons. Above a certain temperature where the thermal energy gets larger than the characteristic trapping energy, we may expect better thermalization of the excitons and the emission spectrum may then be insensitive to the effects of disorder and be governed by the intrinsic exciton-phonon interaction physics. This is exactly what we observe.

\begin{figure}[h]
	\includegraphics[scale=0.25]{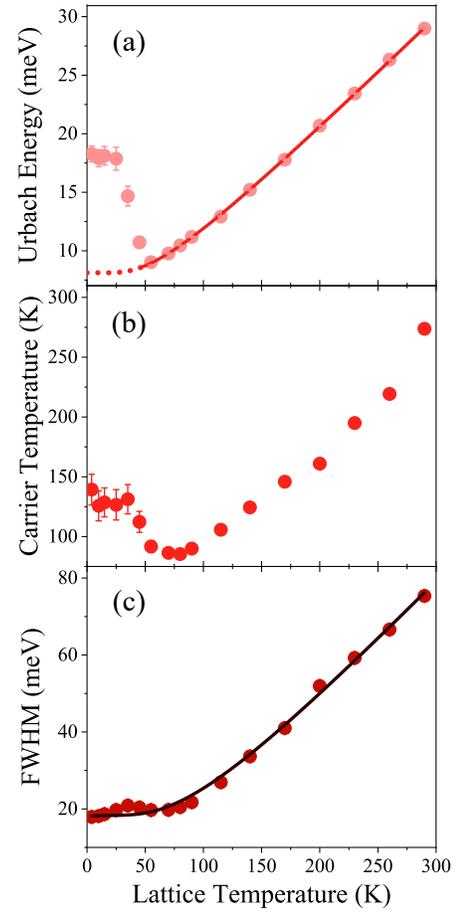}
	\centering
	\caption{(a) Urbach energy $E_U$ as a function of lattice temperature. Solid circles are the data points and the solid line represents a fit to Equation $\ref{Eq:2}$. The fits yield $E_U^0=8.13$ meV and $\hbar\Omega_p=14.3$ meV. The dashed line shows the extrapolation of Urbach energy fit below 55 K. (b) Carrier temperature as a function of lattice temperature. (c) FWHM of the PL peak as a function of lattice temperature. The solid line indicates the fit of Equation $\ref{Eq:7}$ to the experimental data. Here, lattice temperature is assumed to the same as the cryostat temperature.}
	\label{Fig:5}
\end{figure}

In Figure $\ref{Fig:3}$ (a) and (b), we have plotted the inferred absorbance from the PL spectra at different temperatures. The peaks in each case have been arbitrarily scaled to unity, and the abscissa in each case has also been shifted to correct for the anomalous bandgap shift. Following the fit to the temperature dependence of the bandgap [Figure $\ref{Fig:2}$ (b)], the excitonic peaks now appear at the zero of energy. The low energy edge where the value of the absorbance is between $2\times 10^{-1}$ and $2\times 10^{-2}$ are identified as the Urbach tail regions. Figure $\ref{Fig:3}$ (a) shows the spectra around 55 K where we observe a systematic decrease in the slope with increasing temperature. We note that the Urbach rule is clearly established for the spectra above 55 K, viz., (i) we observe an exponential density of states, with (ii) a clearly discernable focus $\varepsilon_F$ at about 33 meV. This value of $\varepsilon_F$ is identified with the zero temperature bandgap renormalization.\cite{Basab_JPhys,Rupak_PRL} Finally, (iii) the temperature dependence of the Urbach slope $E_U(T)$ itself obeys Equation $\ref{Eq:2}$, as can be seen in Figure $\ref{Fig:5}$ (a), with $E_U^0=8.13$ meV and $\hbar\Omega_p=14.3$ meV.
We also note from Figure $\ref{Fig:4}$ (a) and (b) that for temperatures around 70 K, the inferred $E_U$ is invariant over a large range of excitation power, indicating that intrinsic phonon-mediated physics is at play at these temperatures.

In Figure $\ref{Fig:5}$ (b), the carrier temperature $T_X$  is inferred by fitting the high energy tails\cite{Pettinari, Falsini2022} of the free exciton emission peak to an exponential function. The fitting was accomplished by a linear fit to tail of emission spectrum plotted on the semi-log scale, viz., $\log_e[\Im(\hbar\omega)]=A-[{1\over k_BT_X}]\hbar\omega$. The carrier temperature above 70 K nearly follow the lattice temperatures and the inferred carrier temperatures are largely insensitive to the excitation powers [Figure $\ref{Fig:4}$ (c) and (d)].

The temperature-dependent linewidth of the free exciton PL spectra depicted in Figure $\ref{Fig:5}$ (c). Note that the emission spectra at higher temperature are highly asymmetric [Fig $\ref{Fig:2}$ (a)] suggesting another lower energy peak. \cite{Tao2021} To avoid any contribution from these low-energy states, we calculate the ``FWHM" by doubling the high energy half-width-at-half-maximum of the free exciton peak. Note that the shape of the temperature dependence of this linewidth above 50 K closely resembles the temperature dependence of $E_U(T)$, indicating that the high temperature behaviour is determined by the scattering by optical phonons of similar characteristic energy.

More concretely, the temperature dependence of the FWHM $\Gamma(T)$ can be phenomenologically described by the equation\cite{Ramade2018}
\begin{equation}\label{Eq:7}
	\Gamma(T)=\Gamma_0+\sigma_{A}T+\gamma_{LO}\left(e^\frac{\hbar\tilde{\Omega}_p}{k_BT}-1 \right)^{-1}
\end{equation}
Here $\Gamma_0$ is the temperature-independent inhomogeneous broadening due to the scattering from imperfection and the scattering from zero-point phonons. The homogeneous broadening of the emission line is accounted for by considering\cite{Wright2016}: (i) the anharmonic deformation, captured through acoustic phonons (ii) Fr\"ohlich interactions, mediated by the longitudinal optical (LO) phonons. $\sigma_{A}$, $\gamma_{LO}$, $E_{LO}$ in the above equation are the free exciton-acoustic phonon coupling coefficient, exciton-LO phonon coupling strength, and the characteristic LO phonon energy respectively. The solid line in Figure $\ref{Fig:5}$ (c) shows the fit to experimental data with Equation $\ref{Eq:7}$. The estimated parameters are: $\Gamma_0 =$ 17 meV, $\sigma_{A} =$ 2.7 $\times$ 10$^{-4}$ meVK$^{-1}$, $\gamma_{LO} =$ 48 meV, and $\hbar\tilde{\Omega}_p =$ 15.5 meV. The small value of $\sigma_{A}$ suggests a minimal contribution of acoustic phonons to the spectral broadening, and the key role in the scattering process is performed by the LO phonons.\cite{Liu2019} $\hbar\tilde{\Omega}_p=15.5 $ meV agrees well with $\hbar\Omega_p=14.3$ meV  was inferred from the temperature dependence of the Urbach slope $E_U(T)$ in Fig. 5 (c).

The character of the emission spectrum above 70 K is thus unambiguously intrinsic and dominated by the exciton-phonon scattering. The spectral shape, including that of the low energy Urbach tail and the high energy tail (used to infer the carrier temperature), is invariant over a large range of excitation power.

We thus find that (apart from our inability to explain the pronounced asymmetry in the linsehape) the emission spectra at temperatures above 70 K are well-explained assuming excitonic emission from Mott-Wannier excitons under moderate disorder and weak exciton-phonon coupling. While the lineshape is largely invariant above 70 K, the behaviour of the PL spectra below 55 K shows localization anomalies. The Urbach slope, the carrier temperature, and the linewidth all consistently show the same non-monotonic behaviour that may be taken to be the signature of extrinsic effects resulting from defect-mediated localization.  Due to the relatively smaller density of the localized states, we observe a strong excitation power-dependence of both the Urbach slope and the carrier temperature at low temperatures, with the excitation power dependence being the strongest at 4 K [Figure $\ref{Fig:4}$]. Based on these observations, we can associate a characteristic trap energy of $k_BT\approx 5$ meV (using  $T\approx 55 $ K). We emphasize that these low temperature anomalies are also not unusual and have been systematically documented in the context of GaAs quantum wells and nitride semiconductors.\cite{Runge}

In conclusion, we have established the Urbach rule for CsPbBr$_3$ nanosheets, including the demonstration of the Urbach focus. The identification of the Urbach focus has allowed us to infer the zero-temperature bandgap renormalization due to the zero-point phonons' contribution to the exciton-phonon interactions. The value of the renomalization energy at 33 meV is rather modest and rules out strong exciton-phonon interactions. Indeed, the excitons in this material seem to be of the Mott-Wannier type, much like those in materials like GaAs under conditions of weak-to-moderate disorder. They are trapped at low temperature and undergo a gradual localization-to-thermalization (LT) crossover.

See the Supplementary Material for sample synthesis and characterization, details about the PL measurements, importance of emission and absorption experiments in the lights of disorder landscape and Stokes shift, a discussion about whether Dow-Redfield or Toyozawa's theory should be used to explain the electron-phonon coupling in our system, two-dimensional correlation analysis of the emission spectra, and determination of the exciton binding energy.

KM thanks Rupak Bhattacharya for many fruitful suggestions. AM thanks University Grants Commission, New Delhi, for his fellowship. BR thanks Council of Scientific and Industrial Research for his fellowship. SB thanks to the financial support from SERB under Sanction No. CRG/2020/000084 and STR/2021/000001. BB thanks Science and Engineering Research Board, Department of Science and Technology, Government of India, for the Core Research Grant (CRG/2018/003282).

\end{document}


\title{Carrier thermalization and zero-point bandgap renormalization in halide perovskites from the Urbach tails of the emission spectrum}
\author{Kingshuk Mukhuti}\thanks{Authors to whom correspondence should be addressed: kingshukmukhuti.phy@gmail.com and bhavtosh@iiserkol.ac.in}
\affiliation{Department of Physical Sciences, Indian Institute of Science Education and Research Kolkata, Mohanpur, Nadia 741246, West Bengal, India}
\author{Arnab Mandal}
\affiliation{Department of Chemical Sciences and Centre for Advanced Functional Materials, Indian Institute of Science Education and Research Kolkata, Mohanpur, Nadia 741246, West Bengal, India}
\author{Basabendra Roy}
\affiliation{Department of Physical Sciences, Indian Institute of Science Education and Research Kolkata, Mohanpur, Nadia 741246, West Bengal, India}
\author{Sayan Bhattacharyya}
\affiliation{Department of Chemical Sciences and Centre for Advanced Functional Materials, Indian Institute of Science Education and Research Kolkata, Mohanpur, Nadia 741246, West Bengal, India}
\author{Bhavtosh Bansal}\thanks{Authors to whom correspondence should be addressed: kingshukmukhuti.phy@gmail.com and bhavtosh@iiserkol.ac.in}
\affiliation{Department of Physical Sciences, Indian Institute of Science Education and Research Kolkata, Mohanpur, Nadia 741246, West Bengal, India}
\maketitle
\tableofcontents
\section{Synthesis and Characterization}
Synthesis and the characterization of the prepared sample is done according to our previously reported method\cite{Roy2020, Mandal2021} which we discuss briefly in the following section.\\ ~ \\
\textbf{Chemicals:}
Cs$_2$CO$_3$ (99.9\%, Aldrich), octadecene (ODE, 90\%, Aldrich), oleic acid (OA, 90\%, Aldrich), PbBr$_2$ (99.999\%, Aldrich),  oleyl amine (OAm, Aldrich, 70\%), octanoic acid (OctA, 98\%, Aldrich), octylamine (OctAm, 98\%, Spectrochem) and hexane (synthesis grade, Merck) were used without further purification.\\

\textbf{Sample Preparation:}
To prepare the CsPbBr$_3$ nanosheets, 0.4 mmol of Cs$_2$CO$_3$ and 5 ml of OA are loaded in a three-neck flask and dried under vacuum at 120 $^\circ$C for 1h followed by heating at the same temperature at inert atmosphere for 30 min. The Cs-oleate precursor solution is stored under inert condition at room temperature. 5 ml of ODE and 0.19 mmol of PbBr$_2$ are stirred under vacuum for 1h at 120 $^\circ$C followed by N$_2$ purging. Short chain ligands (OctA and OctAm) of 0.4 ml and 0.1 ml of long chain ligands (OA and OAm) are added and kept under the N$_2$ flow till the solution becomes clear. Cs-oleate solution of 0.4 ml is added at 150 $^\circ$C and after 5 min the reaction mixture is introduced into an ice-water bath to quench the reaction.

The crude dispersion of CsPbBr$_3$ nanosheets is centrifuged at 7000 RPM for 15 min. After centrifugation, the supernatant is discarded and the nanosheets are redispersed in 10 ml hexane. 20 ml of methyl acetate is added into the hexane dispersion followed by centrifugation for 10 min at 5000 RPM. The process is repeated twice and the precipitate is kept in a desiccator overnight. In order to prepare CsPbBr$_3$ nanosheets, 30 mg/ml solution of perovskite in chlorobenzene was spin-coated at the top of sapphire slides at 2000 RPM for 30 sec followed by removal of excess ligands in a saturated solution of Pb(NO$_3$)$_2$ in methyl acetate inside the glove box.\\

\begin{figure}[h]
	\includegraphics[scale=0.15]{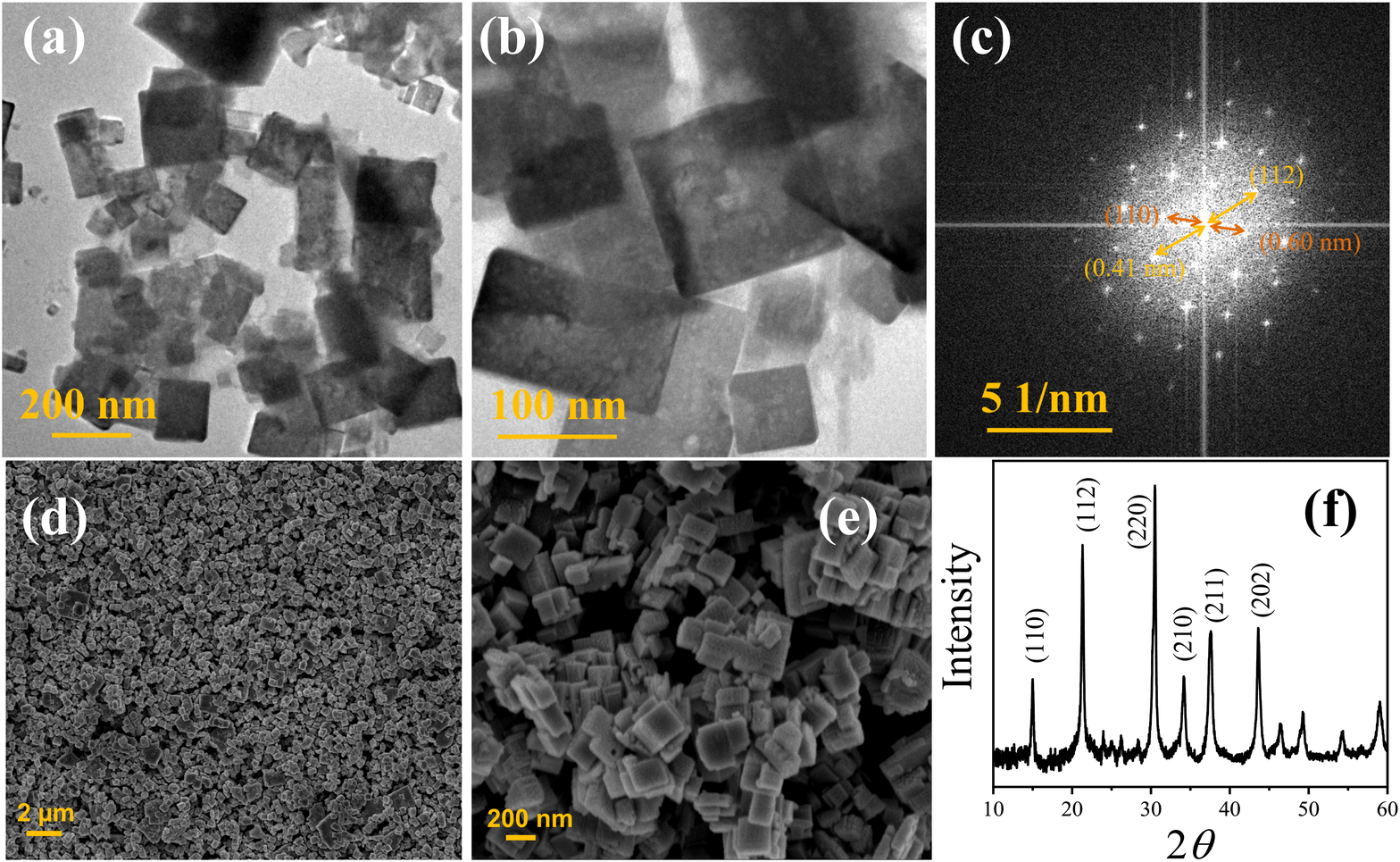}
	\caption{(a), (b) TEM images of CsPbBr$_3$ nanosheets at different magnifications. (c) SAED pattern of the nanosheets. (d), (e) SEM images of the nanosheets at different magnifications. (f) XRD pattern of the nanosheets.}
	\label{Fig:S1}
\end{figure}

\textbf{Structural Analysis:}
Figure $\ref{Fig:S1}$ (a)-(b) represents the TEM images of  CsPbBr$_3$ nanosheets at different magnifications, indicating the formation of rectangular-shaped sheets with lateral size less than 200 nm.
Selected area electron diffraction (SAED) patterns of the nanosheets [Figure $\ref{Fig:S1}$ (c)] indicates the crystalline nature of the nanosheet material. SEM images in Figure $\ref{Fig:S1}$ (d)-(e) also indicate similar results as obtained from TEM images. The crystalline nature of the sheets is further confirmed by the powder XRD pattern [Figure $\ref{Fig:S1}$ (f)], indicating the orthorhombic crystal structure with \textit{Pbnm} space group.\\ 

\textbf{Quantum Yield Measurement:}
Photoluminescence quantum yield (QY) is measured with the help of a standard dye Coumarin 153, and is calculated according to the equation

\begin{equation}
(QY)_s = (QY)_r \frac{\eta_s^2 I_s A_r}{\eta_r^2 I_r A_s}
\label{eq:S1},
\end{equation}
where, (QY)$_s$ and (QY)$_r$ are QY of the sample and reference, respectively. $\eta_s$ and $\eta_r$ are the refractive indices of the solvent used for the sample and the reference, respectively. $I_s$ and $I_r$ are the integrated fluorescence intensity of the sample and reference, respectively. $A_s$ and $A_r$ are the absorbance of the sample and reference at their respective excitation wavelengths, respectively. Due to the quantum confinement effect the CsPbBr$_3$ nanosheets have very high QY, and is calculated to be 74\% using Equation $\ref{eq:S1}$.\\

\section{Photoluminescence Measurement}
The films used in this study were deposited on sapphire substrates and is mounted on the cold finger of a helium closed cycle cryostat, where the sample temperature was varied from 4 K to 290 K. It should be mentioned that thin film samples reduces the self-absorption and luminescence filtering effects which are known to corrupt the PL spectra measured on bulk samples of perovskite materials. For photoluminescence measurements, The sample with a laser light of 403 nm from a CW diode laser, focussed to an incident spot size of $\sim$50 $\mu$m in diameter. Very strong photoluminescence from the sample was collected (with a lens of $f=$10 cm, \o=2 inch) and focused (with a lens of $f=$20 cm, and \o=2 inch to match $f$-number) on to the entrance slit of an Andor spectrograph attached with an electron-multiplying-CCD. The incident laser power was varied over four decades with the help of neutral density filters.

\section{Emission {\em vs} Absorption measurements}
There is a fundamental difference on the account of the Urbach rule being inferred from the emission, rather than the absorption spectrum. This work tries to build on this difference and so we further elaborate it here.

The absorption spectrum is proportional to the joint density of states in the valence band or relevant to this work, the density of states of the excitons. The emission spectrum on the other hand maps the distribution of the excitonic states as they are occupied under the given conditions of temperature and chemical potential (the excitation density).\cite{Runge}

\subsection{Disorder landscape}
On the account of spatial disorder in any sample, the bandgap is not constant in space but varies from point to point, thereby forming a potential ``landscape."  This landscape, viz, the location of the bandedge as a function of the spatial coordinates, is characterized by a typical energy scale, $\Delta$. $\Delta$ is height of a typical barrier between two typical ``valleys". It is then easy to see that the carrier dynamics is separated into two extreme limits, (i) when $\Delta >> k_BT$, the carriers don't have sufficient energy to escape the local trapping site and, basically, stay trapped in the first local minimum they diffuse into, until they recombine. The PL spectrum reflects the density of states of the minima of the potential landscape in this case.  (ii) When $\Delta << k_BT$, the carriers have sufficient energy to overcome the energy barriers and transition between the valleys. The carriers can then be expected to equilibrate to a Boltzmann-like distribution within the recombination time scale of a few nanoseconds. For such thermalized carriers, the structure of the landscape is not particularly important in determining the PL lineshape. We would thus expect the PL characteristics to be strongly dependent on sample specific extrinsic disorder at low temperatures but be determined by intrinsic features (phonons and the equilibrium carrier distribution) at higher temperatures.
From figure 5 [main text],  $\Delta/k_B$ to be about 75 K in our experiments because the signatures of thermalization can be seen beyond this temperature.
A detailed and rigorous discussion of these issues may be found in the review by E. Runge.\cite{Runge}

Also note that under the usual conditions of excitation, the number of carriers participating in the PL emission is orders of magnitude smaller than the total number of states in the band and that the PL spectrum preferentially reflects the small number of low energy states. At low temperature, these are the local minima of the potential landscape. As the laser excitation density is increased, many of the localized minima may already be filled and thus be not available to the carriers. One therefore sees an effectively smoother landscape at higher excitation powers and the carriers now have a better  chance of thermalizing. This is exactly what is seen in Figure 4 [main text]. At low temperatures, there is a strong dependence of the Urbach energy with power whereas at higher temperature, when the carriers are already thermalized, the parameters are nearly excitation power-independent.
\subsection{Stokes shift}
\begin{figure}[h]
	\includegraphics[scale=0.28]{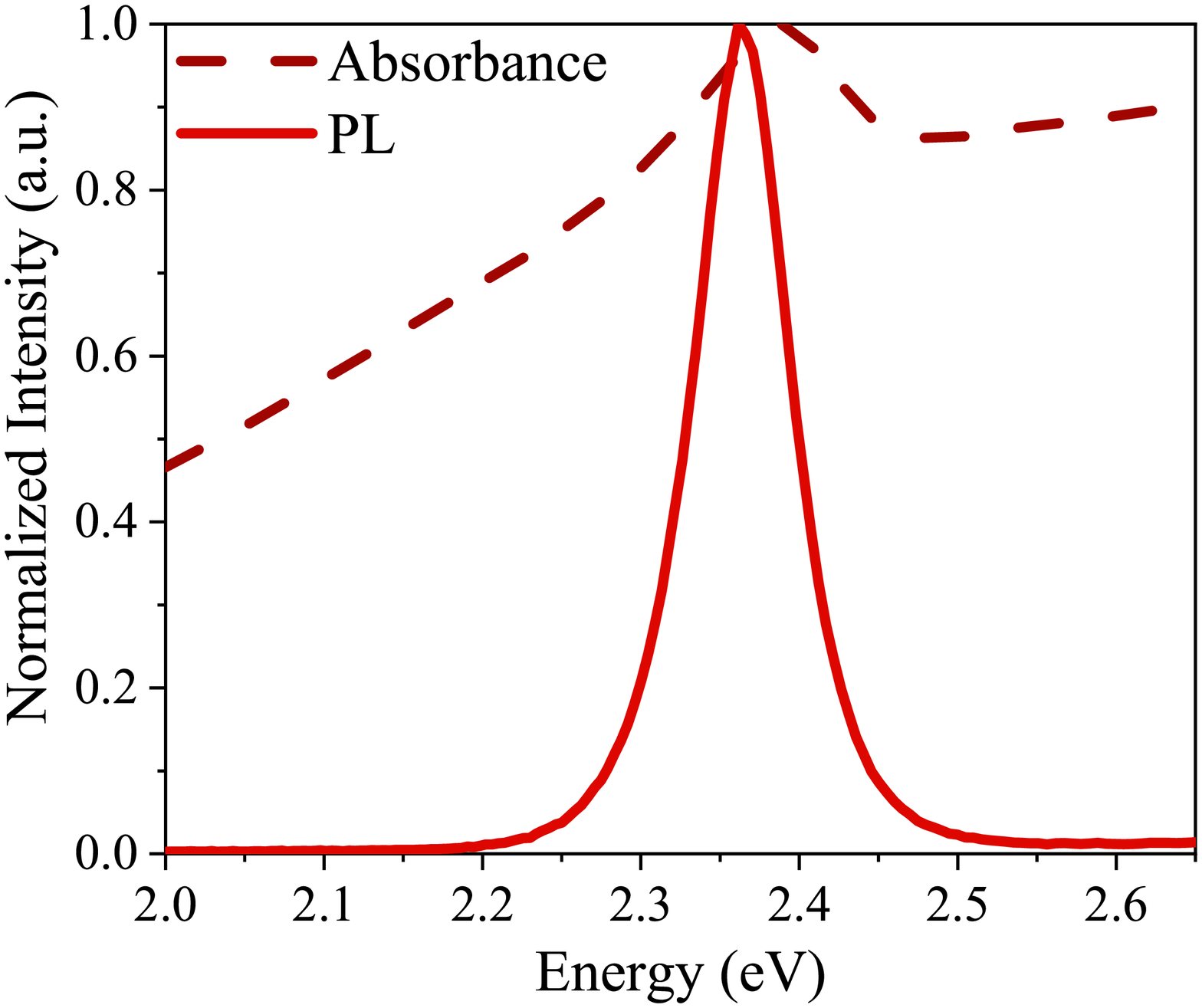}
	\centering
	\caption{Absorption and PL emission spectra of the nanosheets at room temperature. There is a Stokes shift of about 25 meV between the emission and the absorption spectrum at room temperature.}
	\label{Fig:S2}
\end{figure}

Figure \ref{Fig:S2} shows that the absorption and emission spectra at room temperature have a Stokes shift of about 25 meV. This agrees with other published reports.\cite{Brennan} The observed Stokes shift is also consistent with what may be expected for thermalized carriers;\cite{Runge} when the excitonic line has a Gaussian lineshape, elementary arguments suggest that the Stokes shift $\Sigma$ should be related to the absorption linewidth $\Delta$ via the relation $\Sigma\approx 0.18\Delta^2/k_BT_e$.\cite{Pettinari} For $\Delta=76$ meV, this relation predicts $\Sigma=40$ meV for $T=300$ K.

\section{Dow-Redfield {\em vs} Toyozawa theory}
One of the active points of discussion in various halide perovskites has been the role of exciton-phonon interactions, with many reports suggesting manifestations of polaron physics in the PL spectra.\cite{Shuran2019, Kandada2020} As has been indicated above, we have found no evidence of strong exciton-phonon interactions and the theory for Mott-Wannier excitons under weak-coupling and moderate disorder explains all the observations (except the asymmetry in the lineshape).

The Urbach rule  [Equation 1 and 2 in the main paper] for the excitonic absorption edge is described by two competing theories, both invoking the exciton-phonon interaction. Since both the theories have the same specific predictions, all our conclusions including the inference of the zero-point renomalization are model-independent.

Dow and Redfield's theory explains the Urbach edge as being due to the perturbation on the hydyogenic excitons on account of the effective electric field that they experience (in the relative coordinate frame of orbiting electrons and holes) from the polar optical phonons. This electric field leads to a finite tunnelling probability for the ionization of the bound state and the Urbach edge is to be treated as a special case of this lifetime broadening.\cite{Dow-Redfield,Schafer-Wegener, Rupak_PRL, Haug_Schmitt-Rink} This theory is generally believed to be applicable to the weak coupling Mott-Wannier excitons.

The rival theory had been developed by Toyozawa and many coworkers over the span of three decades.\cite{Toyozawa_Book, Sumi1987, Song-Williams} This theory is usually thought to be relevant under conditions of stronger exciton-phonon coupling, as is the case for alkali halides. The Urbach tails are derived as the consequence of the acoustic phonon-mediated short-ranged interaction affecting the excitons' {\em centre of mass} coordinate.\cite{Tao2021, Schilcher2021, Sumi1987, Rashba} While the Urbach rule is derived from a complex and opaque calculation (viz., involving heavy numerical work), roughly speaking, the polaronic coupling can be visualized as leading to defect-like bound states for the excitons.\cite{Toyozawa_Book} The Urbach tail can be thought as being formed on account of the trapping-detrapping of the excitonic polarons in and out of these dynamically formed bound states.

Since the basic predictions of the two theories [Equation 1 and 2 in the main paper] are the same, it has been hard to falsify one in favour of the other. While Sumi\cite{Sumi1987} has passionately argued in favour of Toyozawa, an experimental study of electroabsorption in CuCl and TlCl favours the Dow-Redfield model.\cite{Mohler-Thomas} In the context of halide perovskites, the observed asymmetry in the lineshape of the emission spectrum at room temperature (the phenomenon we could not explain) has been interpreted as a secondary low energy bound-polaron peak and the observed Urbach tail has been attributed to these polaronic effects in the intermediate-coupling trapping-detrapping regime.\cite{Tao2021, Tao2022}

The strength of the exciton-phonon coupling using Toyozawa theory can be estimated from the value of the slope parameter $\sigma (T)$. Here $\sigma (T)$ is defined by writing Equation 1 [main paper] in equivalent form,\cite{Toyozawa_Book, Song-Williams} viz., $\alpha(\hbar\omega,T)=\alpha_{0}\exp\left[{\sigma(T)\over k_BT}{(\hbar \omega-\varepsilon_F)}\right]$ with $\sigma (T)={ 2 k_BT\over \hbar\Omega_p}\sigma_0 \tanh({\hbar\Omega_p\over 2k_BT})$. From Equation 2 [main paper], given our values of $E_U^0=8.13$ meV and $\hbar\Omega_p=14.3$ meV, we have $\sigma_0=[{2E_U^0\over \hbar\Omega_p}]^{-1}=0.88$.
\begin{figure*}[t]
	\includegraphics[scale=0.19]{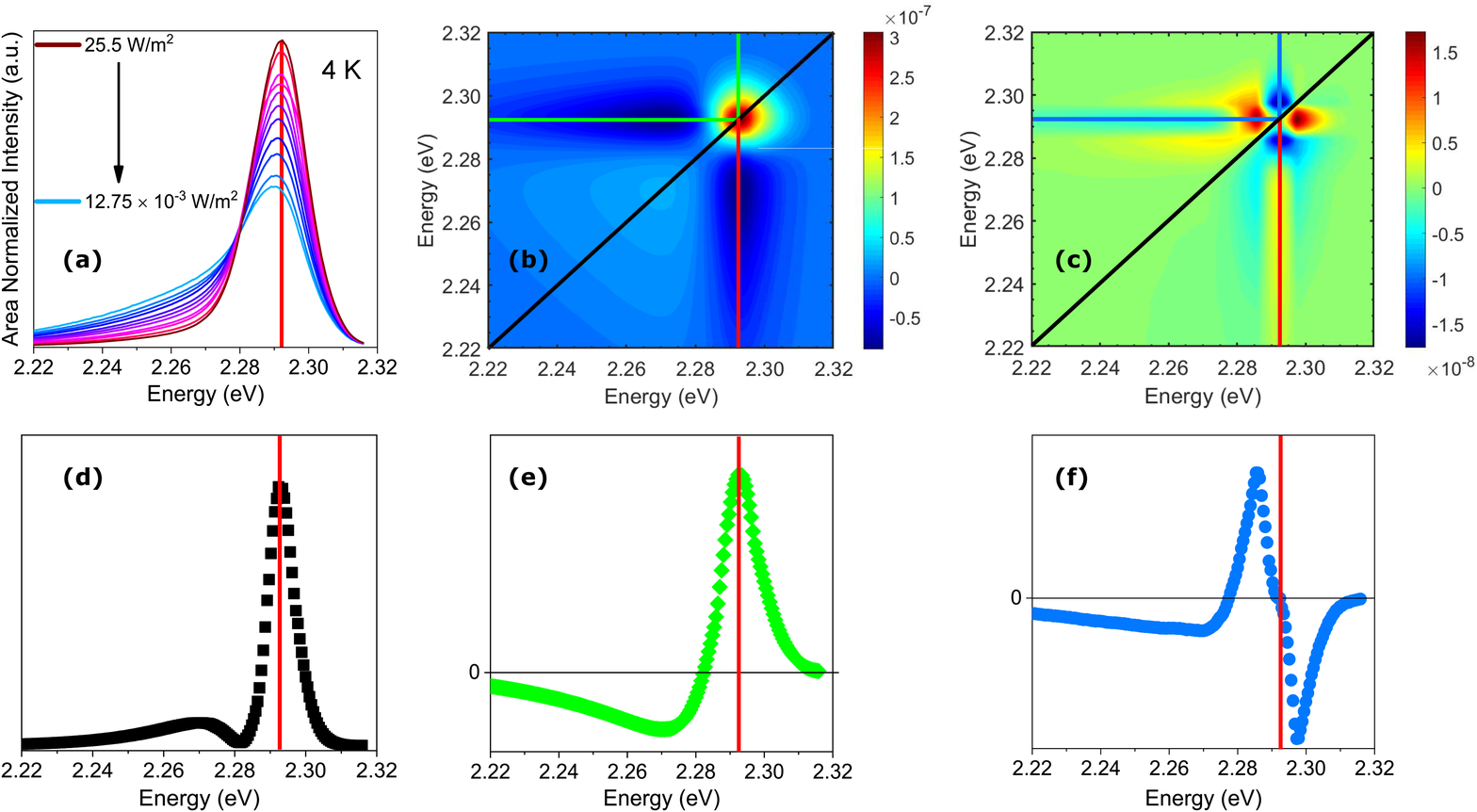}
	\centering
	\caption{The vertical lines mark the position of the free exciton peak. (a) Area-normalized photoluminescence spectra for varying excitation density measured at 4 K. (b) Synchronous correlation diagram calculated from the excitation-dependent spectra. Colors represent the strength of the correlation. This shows the `in-phase' correlated intensity-variation of different energies on the spectra with respect to the change in excitation density. (c) Calculated asynchronous correlation diagram captures the `out of phase' intensity-variation at energies on the spectra that are changing at different rates with varying excitation density. (d) Synchronous spectrum on the diagonal black line. This clearly separates the free and the localized excitons on the energy axis. The position of maximum `autopeak' matches quite well with the free exciton peak. (e) Intensity-variation at different energies with respect to the maximum autopeak as a function of excitation density, depicted on the green line in synchronous spectrum. (f) Rate of change of intensity-variation at different energies with respect to the maximum autopeak as a function of excitation density, depicted on the red line in asynchronous spectrum.}
	\label{Fig:S3}
\end{figure*}
$\sigma_0$ is inversely related to the strength of the exciton-phonon interaction $g=E_{LR}/B$, where $E_{LR}$ is the lattice relaxation energy and $B$ is the excitonic bandwidth. Following Toyozawa, $g=s/\sigma_0$, where $s$ is the called the steepness index, dependent only on the crystal structure and its dimension; for a three-dimensional cubic lattice $s\approx 1.5$ and $s\approx 1.24$ for a two-dimensional square lattice.\cite{Toyozawa_Book, Song-Williams} The critical coupling strength separating the weak from the strong coupling regimes is $g=0.92$ for a simple cubic lattice and $g=0.088$ for a two-dimensional square lattice. Our estimate for $\sigma_0=0.88$ yields $g\approx 1.4$ (assuming two dimensional sample) indicating that the value of the electron-phonon coupling does indeed exceed the critical threshold for the formation of the bound state within the Toyozawa theory. The Toyozawa theory would thus indicate intermediate to strong polaronic coupling. But this fact does not quite seem to be borne out by other observations. In particular, we observe no phonon replicas.

A strong coupling would also imply the temperature dependence of the linewidth $\Gamma(T)$ scaling as $\Gamma(T)\sim \sqrt{\coth{{\hbar \omega_p\over k_BT}}}$ whereas a $\Gamma(T)\sim \coth{{\hbar \omega_p\over k_BT}}$, i.e., the linewidth scaling with the phonon population is what is observed. Thus the Dow-Redfield model seems to be better describing our observations.

\section{Two-dimensional Correlation Analysis}
We have noted that the PL spectra at low temperatures were strongly power-dependent with an indication of a bound state at 2.275 eV. The existence of this bound state can be systematically studied using the two-dimensional correlation analysis on the excitation power-dependent spectra. Pioneered by Noda and coworkers\cite{Noda2005}, the main idea is to first get a series of spectra [shown in Figure $\ref{Fig:S3}$ (a)] with respect to an external variable (happens to be excitation power in this case). Then to calculate the differences from a suitably chosen spectrum of zero mean. At last, to calculate a correlation among these differences at various energies. The correlation matrix is then expressed in terms of two orthogonal matrices, namely \textit{synchronous} and \textit{asynchronous} correlation intensities. The algorithm is fairly straightforward and is explained elsewhere \cite{Noda2005, Mukhuti2019}.

By construction, the calculated synchronous correlation diagram in Figure $\ref{Fig:S3}$ (b) is a symmetric matrix and the colors represent the strength of the correlation of a spectrum with itself. The diagonal black line trivially denotes that all the energies are perfectly correlated with itself. Thus any correlation patch that appears on the diagonal is called an autopeak and notably, the strongest autopeak appears at the position of the free exciton peak. When plotted with respect to energy, in a sense it deconvolutes the embedded features of the original spectra and we plot the same in Figure $\ref{Fig:S3}$ (d). The vertical red line marks the position of the free exciton peak and match remarkably well in Figures $\ref{Fig:S3}$ (b) and (d). Also, the bound excitonic features are now visible both in the synchronous correlation diagram [Figure $\ref{Fig:S3}$ (b)] and in the diagonal plot [Figure $\ref{Fig:S3}$ (d)] at the low energy side of the free exciton peak. Alternatively, one could try to fit different Gaussian envelopes to the spectra and figure out their contribution. However, two-dimensional correlation analysis provides significantly more information than a mere curve-fitting protocol. The off-diagonal patches in Figure $\ref{Fig:S3}$ (b) captures those energies where the correlation intensity changes together or one at the expense of the other depending upon whether the patch appears in the same or in the opposite color. With respect to the strongest autopeak, the synchronous correlation intensities at other energies is pointed out by the green line on Figure $\ref{Fig:S3}$ (b), and is presented in Figure $\ref{Fig:S3}$ (e). Thus on an area normalized scale, as the excitation power is increased, the free excitons grow at the expense of the bound ones. Therefore, on a normal scale this would mean that with excitation density, both the free and the bound exciton emission is increasing. However, the rate of increase in free exciton emission is much higher than the bound excitons. Clearly, the PL signal at low temperature is

The asynchronous correlation diagram in Figure $\ref{Fig:S3}$ (c) is an antisymmetric matrix and in close comparison to the analytical signals, it is the calculated correlation between a spectrum and its Hilbert transform with respect to excitation density. A nonzero asynchronous signal thus appears only when there is a `out of phase' intensity variation at two energies with respect to excitation density. Therefore, it is trivially satisfied that there is no asynchronous correlation signal on the diagonal (black line) in Figure $\ref{Fig:S3}$ (c). Loosely speaking, it is a correlation between a quantity and its rate of change. The asynchronous correlation signal with respect to the free exciton peak is portrayed with the blue line in Figure $\ref{Fig:S3}$ (c). When it is plotted in Figure $\ref{Fig:S3}$ (f) with energy, one finds the relative variation of the bound and the tail states.

\section{Exciton Binding Energy}
\begin{figure}[h]
	\includegraphics[scale=0.26]{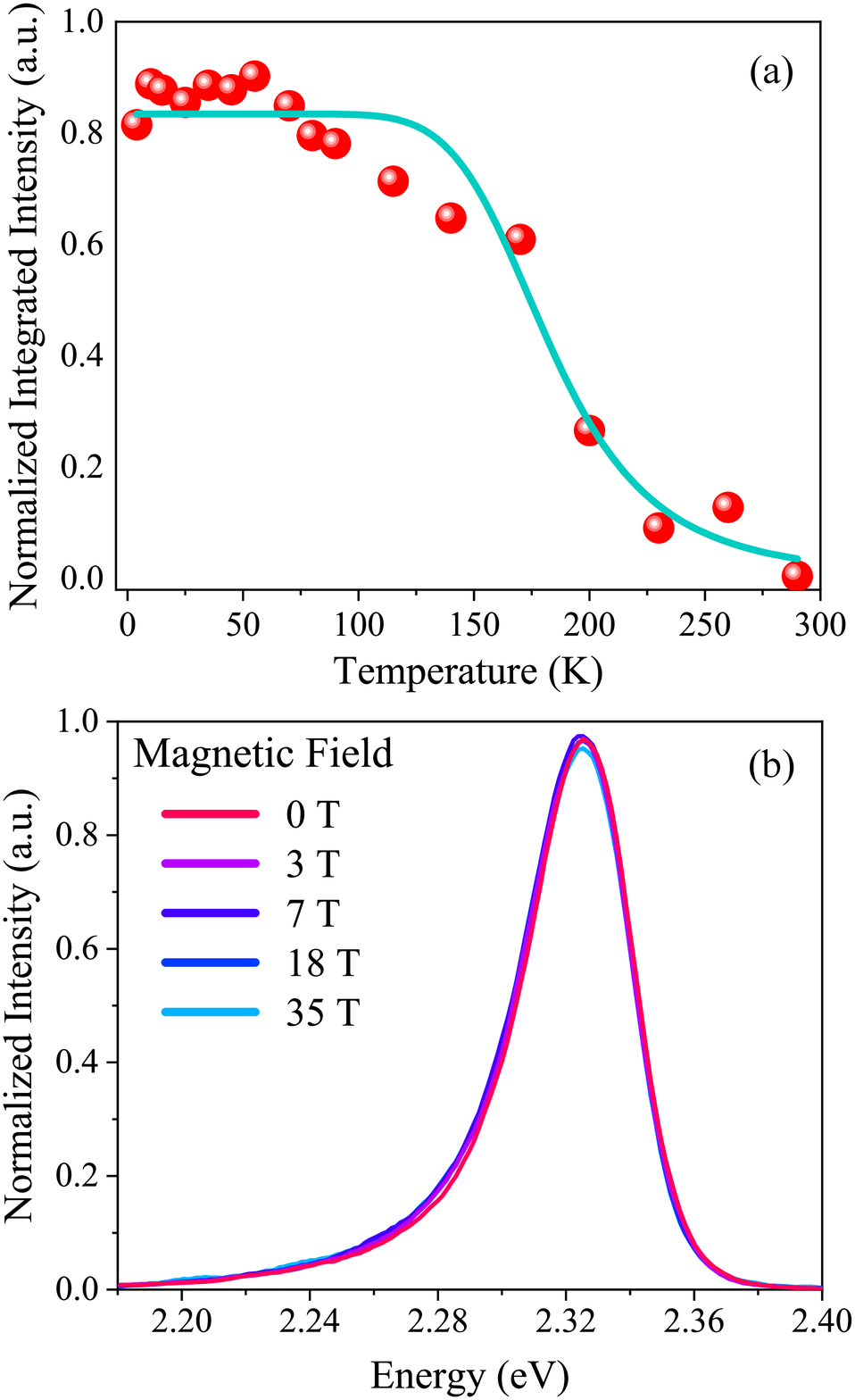}
	\caption{(a) Normalized integrated intensities as a function of temperature. The filled circles represent the data points, and solid line indicates the fit to Equation $\ref{eq:S2}$. (b) Magneto-PL data acquired at 80 K. There is no detectable diamagnetic shift in the free exciton peak even under a magnetic field of 35 T.}
	\label{Fig:S4}
\end{figure}
Unlike the conventional semiconductors (such as GaAs, CdSe, InP), the perovskite nanosheets exhibit bright excitonic luminescence even at ambient conditions. Thus, one would naively guess the exciton binding energy $E_b$ is substantially higher than the thermal energy at room temperature ($\ie$ $E_b\gg 26$ meV). It has been shown that one can calculate the free exciton binding energy from the temperature dependence of the integrated free exciton PL intensity \cite{He2016}, viz.,
\begin{equation}
	I_F(T)=\frac{I_0}{1+\alpha T^{3/2}+e^{-\frac{E_b}{k_BT}}}
	\label{eq:S2}.
\end{equation}

$I_0$ is the intensity at low temperature, $E_b$ is activation or exciton binding energy, and $\alpha$ is a constant.

This equation, when fitted to the normalized integrated intensities originating from the free exciton photoluminescence [Figure $\ref{Fig:S4}$ (a)], yields a large binding energy of 103 meV which is same as the difference between the free exciton peak to the band-edge in Figure 1 [main paper].

Considering the 2D nature of excitons in the CsPbBr$_3$ nanosheets, such values are quite reasonable and matches with previous reports. To further consolidate the results, we perform magneto-photoluminescence measurements on the nanosheets to directly measure the exciton binding energy from the diamagnetic shift of the free exciton photoluminescence peak. One can observe from the Figure $\ref{Fig:S4}$ (b) that the diamagnetic shift in the spectrum is less than our instrumental resolution even at 35 T. Following the footsteps of MacDonald and Ritchie, we can set a satisfactory lower limit of 98 meV for the binding energy, very close to our estimated value.